\documentstyle[12pt]{article}

\makeatletter \@addtoreset{equation}{section} \makeatother

\begin{document}

\begin{titlepage}

\begin{flushright}
FIAN/TD/03/03\\
January 03\\
\end{flushright}

\vspace{3cm}

\begin{center}
{\Large\bf BRST-Invariant Constraint Algebra}
\end{center}

\begin{center}
{\Large\bf in Terms of Commutators and Quantum Antibrackets}
\end{center}

\vspace{1cm}

\begin{center}
{\large Igor Batalin$^{\,\dag}$ and Igor Tyutin$^{\,\ddag}$ }
\end{center}
\begin{center}
{\em Lebedev Physical Institute RAS,\\
53, Leninsky Prospect, 117924 Moscow, Russia}
\end{center}

\vspace{1.7cm}

\begin{abstract}
General structure of BRST-invariant constraint algebra is established, in
its commutator and antibracket forms, by means of formulation of
algebra-generating equations in yet more extended phase space. New
ghost-type variables behave as fields and antifields with respect to quantum
antibrackets. Explicit form of BRST-invariant gauge algebra is given in
detail for rank-one theories with Weyl- and Wick- ordered ghost sector. A
gauge-fixed unitarizing Hamiltonian is constructed, and the formalism is
shown to be physically equivalent to the standard BRST-BFV approach.
\end{abstract}

\begin{quote}

\vfill \hrule width 5.cm \vskip 2.mm $^\dag$ {\small \noindent
E-mail: batalin@lpi.ru}

$^\ddag$ {\small \noindent E-mail: tyutin@lpi.ru}

\end{quote}

\end{titlepage}

\vspace{2cm}

\newpage

\section{Introduction}

When quantizing general gauge theories, a basic principle \cite{BecRouSto74}
- \cite{HenTei92} is to construct a BRST-charge Fermionic operator $\Omega$
which satisfies the standard BRST algebra $\Omega^{2}\equiv\frac{1}{2}\left[
\Omega ,\Omega\right]  =0$, $\left[  G_{C},\Omega\right]  =i\hbar\Omega$,
with $G_{C}$ being a ghost number Bosonic operator. As expanded in power
series in ghost canonical pairs
$(C^{\alpha},\bar{\cal{P}}_{\alpha})$, the operator $\Omega$
begins with $\Omega=C^{\alpha}T_{\alpha}+$ more, where $T_{\alpha}$ are
original first-class constraints working as gauge algebra generators.

It is remarkable that there exist BRST-invariant modified constraints
${\cal T}_{\alpha}\!\!=\!(i\hbar)^{-1}\!\!
\left[\Omega,\bar{{\cal P}}_{\alpha}\right]$ $=T_{\alpha}+$more, satisfying
$\left[\Omega,{\cal T}_{\alpha}\right]=0$ by construction, which determine
substantially a new, dynamically prescribed, set of gauge algebra generators.
As these generators depend actually on ghost operators $(C^{\alpha},
\bar{{\cal P}}_{\alpha})$, they live in an extended phase space,
in contrast to original first-class constraints $T_{\alpha}$.

The main idea of the present paper is to reformulate the standard BRST-BFV
quantization scheme directly in terms of BRST-invariant constraints
${\cal T}_{\alpha}$ considered as new basic ingredients, by means of further
extension of the phase space previously spanned by original phase variables
and ordinary ghosts $(C^{\alpha},\bar{{\cal P}}_{\alpha})$.

Our main motivation is that new gauge generators ${\cal T}_{\alpha}$ are
expected to have, in general, certainly better algebraic properties as
compared with original constraints $T_{\alpha}$. As a particular case, we can
mention the well-known situation in Bosonic string theory (see
\cite{GreShwWit87} and references therein) where the algebra of original
Virasoro generators is centrally extended, while the algebra of the
corresponding BRST-invariant generators coincides exactly with the classical
one.

It is a characteristic property of BRST-invariant generators
${\cal T}_{\alpha}$ that their algebra is closed only if original constraints
form a Lie-type algebra with constant structure coefficients. It appears,
however, that the algebra spanned by generators ${\cal T}_{\alpha}$ and ghost
momenta operators $\bar{{\cal P}}_{\alpha}$ is always closed by construction.

The above circumstance allows us to formulate closed generating equations of
the BRST-invariant gauge algebra in further-extended phase space. With this
purpose, we introduce two sets of new ghost-type canonical pairs,
$(B^{\alpha},\Pi_{\alpha})$ and $(B_{\alpha}^*,\Pi_*^{\alpha})$, which
behave as fields and antifields with respect to quantum antibrackets.

It then appears possible to recast the new generating equations of the
BRST-invariant gauge algebra to the form of operator-valued master equation
\cite{BatMar98a} - \cite{BatMar00} formulated in terms of quantum
antibrackets defined originally in \cite{BatMar98c}. In a natural way, these
quantum antibrackets generate operator-valued anticanonical transformations.
We represent their general form and the transformation properties of the
antibrackets by means of ordinary differential equations in an auxiliary
variable.

As a result, we obtain two dual descriptions of BRST-invariant gauge algebra,
in terms of standard commutators and quantum antibrackets. To illustrate the
dualism in technical respect, we consider in more detail the case of a
rank-one gauge theory with Weyl- and Wick- ordered ghost sector.

As usual, $\varepsilon(f)\equiv\varepsilon_{f}$ denotes the Grassmann parity
of a quantity $f$, while $[f,g]$ stands for the standard supercommutator
$[f,g]\equiv fg-(-1)^{\varepsilon_{f}\varepsilon_{g}}gf$ of any two operators
$f$ and $g$. It satisfies the standard Leibniz rule,
$[fg,h]=f[g,h]+[f,h]g(-1)^{\varepsilon_{g}\varepsilon_{h}}$, and Jacobi
identity, $[f,[g,h]]$ $(-1)^{\varepsilon_{f}\varepsilon_{h}}+$cycle$(f,g,h)=0$%
. Other notation is clear from the context.

\section{Quantum antibrackets and anticanonical transformations}

Here we recall main definitions and properties of quantum antibrackets as they
formulated in \cite{BatMar98c} - \cite{BatMar99c}. Then we define
operator-valued anticanonical transformations and derive how quantum
antibrackets behave under anticanonical transformations of their entries.

Let $Q$ be a Fermionic nilpotent operator,
\begin{equation}
\varepsilon(Q)=1,\;Q^{2}\equiv\frac{1}{2}[Q,Q]=0.\label{1.1}%
\end{equation}

Then the general quantum antibracket is defined by the formula
\begin{equation}
(f,g)_{Q}\equiv\frac{1}{2}\left(  [f,[Q,g]]-[g,[Q,f]](-1)^{(\varepsilon
_{f}+1)(\varepsilon_{g}+1)}\right) \label{1.2}%
\end{equation}
for any two operators $f$ and $g$.

It satisfies
\begin{equation}
\varepsilon\left(  (f,g)_{Q}\right)  =\varepsilon_{f}+\varepsilon
_{g}+1,\label{1.3}%
\end{equation}
and
\begin{equation}
(f,g)_{Q}=-(g,f)_{Q}(-1)^{(\varepsilon_{f}+1)(\varepsilon_{g}+1)}.\label{1.4}%
\end{equation}
It follows from (\ref{1.2}) that the modified Leibniz rule
\[
(fg,h)_{Q}-f(g,h)_{Q}-(f,h)_{Q}g(-1)^{\varepsilon_{g}(\varepsilon_{h}+1)}=
\]%
\begin{equation}
=\frac{1}{2}\left(  [f,h][g,Q](-1)^{\varepsilon_{h}(\varepsilon_{g}%
+1)}+[f,Q][g,h](-1)^{\varepsilon_{g}}\right) \label{1.5}%
\end{equation}
and Jacobi identity
\begin{equation}
\left(  f,(g,h)_{Q}\right)  _{Q}(-1)^{(\varepsilon_{f}+1)(\varepsilon_{h}%
+1)}+\hbox{cycle}(f,g,h)=-\frac{1}{2}\left[(f,g,h)_{Q}
(-1)^{(\varepsilon_{f}+1)(\varepsilon_{h}+1)},Q\right]  ,\label{1.6}%
\end{equation}
hold, where
\begin{eqnarray}
(f,g,h)_{Q}\equiv\frac{1}{3}(-1)^{(\varepsilon_{f}+1)(\varepsilon_{h}%
+1)}\left(  \left[  (f,g)_{Q},h\right]  (-1)^{\varepsilon_{h}+(\varepsilon
_{f}+1)(\varepsilon_{h}+1)}+\hbox{cycle}(f,g,h)\right)  =\nonumber\\
=\frac{1}{3}(-1)^{(\varepsilon_{f}+1)(\varepsilon_{h}+1)}\left(  \left[
f,(g,h)_{Q}\right]  (-1)^{\varepsilon_{g}+\varepsilon_{f}(\varepsilon
_{h}+1)}+\hbox{cycle}(f,g,h)\right)\phantom{\,aa} \label{1.7}%
\end{eqnarray}
defines the next, 3-antibracket, for any operators $f$, $g$, $h$.

In its own turn, the 3-antibracket (\ref{1.7}) satisfies the next Jacoby
identity involving the next, 4-antibracket, and so on. In \cite{BatMar99b},
\cite{BatMar99c}, this hierarchy of subsequent higher-order quantum
antibrackets is determined substancially by means of the corresponding
generating mechanism.

Let $B$ be a Bosonic operator, and $A$ is an arbitrary one. Then, it follows
from (\ref{1.6}) that
\begin{equation}
\left(  B,(B,A)_{Q}\right)  _{Q}=\frac{1}{2}\left(  (B,B)_{Q},A\right)
_{Q}-\frac{1}{4}\left[  (B,B,A)_{Q},Q\right]  ,\label{1.8}%
\end{equation}%
\begin{equation}
(B,B,A)_{Q}=\frac{1}{3}\left(  -\left[  A,(B,B)_{Q}\right]  +2\left[
(A,B)_{Q},B\right]  \right)  ,\label{1.9}%
\end{equation}%
\begin{equation}
\left(  B,(B,B)_{Q}\right)  _{Q}=\frac{1}{6}\left[  [B,(B,B)_{Q}],Q\right]
.\label{1.10}%
\end{equation}
Another important consequence of the definition (\ref{1.2}) and nilpotence
condition (\ref{1.1}) reads
\begin{equation}
\left[  Q,(f,g)_{Q}\right]  =[[Q,f],[Q,g]].\label{1.11}%
\end{equation}

Now, let us define an operator-valued anticanonical transformation as follows.
Let $A_{0}$ be an initial operator, and $X$, $\varepsilon(X)=1$, be a
Fermionic anticanonical generator. Then, the equation
\begin{equation}
\frac{dA}{d\lambda}=(X,A)_{Q},\;\left.  A\right|  _{\lambda=0}=A_{0}%
,\label{1.12}%
\end{equation}
determines the anticanonical transformation $A_{0}\rightarrow A$.

It follows from (\ref{1.11}), (\ref{1.12}) that
\begin{equation}
\frac{d}{d\lambda}[Q,A]=[[Q,X],[Q,A]],\label{1.13}%
\end{equation}
and thereby $[Q,X]$ serves as a canonical Bosonic generator transforming
$[Q,A_{0}]$. The general solution to the parametric differential equation
(\ref{1.12}) reads
\begin{equation}
A=\tilde{A}+[Q,Y],\label{1.14}%
\end{equation}
where
\begin{equation}
\tilde{A}=\tilde{A}(\lambda)=e^{\lambda\lbrack Q,X]}A_{0}e^{-\lambda\lbrack
Q,X]},\label{1.15}%
\end{equation}
and
\begin{equation}
Y=-\frac{1}{2}\int_{0}^{\lambda}d\lambda^{\prime}e^{\frac{1}{2}(\lambda
-\lambda^{\prime})[Q,X]}[X,\tilde{A}(\lambda^{\prime})]e^{-\frac{1}{2}%
(\lambda-\lambda^{\prime})[Q,X]}.\label{1.16}%
\end{equation}
There exists a nice interpretation of the solution (\ref{1.14}): the first
term in r.h.s., $\tilde{A}$, is just a canonical transform of $A_0$
with $[Q,X]$ being a generator, while the second term, $[Q,Y]$, is an
``exact'' form. When taking the formula (\ref{1.14}) in the first order in
$X$,
\begin{equation} A-A_{0}=\lambda\left(  \lbrack\lbrack
Q,X],A_{0}]-\frac{1}{2}[Q,[X,A]]\right)
+O(\lambda^{2})=\lambda(X,A_{0})_{Q}+O(\lambda^{2})\label{1.17}%
\end{equation}
we see that the part $[[Q,X],A_{0}]$ of quantum antibracket $(X,A_{0})_{Q}$ in
(\ref{1.17}) is just an infinitesimal transform of $A_{0}$ with $[Q,X]$
being a generator, while the part $-\frac{1}{2}[Q,[X,A]]$ (the ``exact''
one) represents the nonunimodularity characteristic to anticanonical
transformations.

Now, let $A$ and $B$ be anticanonical transforms of $A_{0}$ and $B_{0}$ in
the sense of the equation (\ref{1.12}). Then the quantum antibracket
$(A,B)_{Q}$ satisfies the equation
\begin{equation}
\frac{d}{d\lambda}(A,B)_{Q}=(X,(A,B)_{Q})_{Q}+\frac{1}{2}[(X,A,B)_{Q}%
,Q],\label{1.18}%
\end{equation}
where the modified Jacobi identity (\ref{1.6}) for $f=X$, $g=A$, $h=B$ is
taken into account. It follows from (\ref{1.18}) that the deviation of the
antibracket $(A,B)_{Q}$ from its anticanonical invariance (solution to the
homogeneous part of eq. (\ref{1.18})) is given by the ``exact'' form
\begin{equation} \left[
Q,\frac{1}{2}\int_{0}^{\lambda}d\lambda^{\prime}e^{\frac{1}{2}%
(\lambda-\lambda^{\prime})[Q,X]}(X,A,B)_{Q}e^{-\frac{1}{2}(\lambda
-\lambda^{\prime})[Q,X]}\right]  .\label{1.19}%
\end{equation}

Thus we conclude that the appearance of nonzero ``exact'' form $[(f,g,h)_{Q}%
,Q]$, which deviates the modified Jacobi identity from being a strong one,
results in a similar deviation of the invariance property of quantum
antibracket under anticanonical transformation of its entries.

\section{BRST-invariant constraint algebra}

Let $\Omega$ be a Fermionic operator which satisfies the standard BRST
algebra
\begin{equation} \left[  \Omega,\Omega\right]  =0,\;\left[
G_{C},\Omega\right]  =i\hbar \Omega,\label{2.1}
\end{equation}
with $G_{C}$ being a ghost number Bosonic operator.

For the sake of definiteness, we assume that ghost sector is represented by
canonical pairs $(C^{\alpha},\bar{{\cal P}}_{\alpha})$,
$\varepsilon(C^{\alpha})=\varepsilon(\bar{{\cal P}}_{\alpha})
=\varepsilon_{\alpha}+1$, with the only nonzero commutators
\begin{equation}
\left[  C^{\alpha},\bar{{\cal P}}_{\beta}\right]  =i\hbar\delta_{\beta
}^{\alpha},\label{2.2}%
\end{equation}
and the BRST operator $\Omega$ is $C\bar{{\cal P}}$-ordered. As for the
ghost number assignment, we assume that
\begin{equation}
\lbrack G_{C},C^{\alpha}]=i\hbar C^{\alpha},\;[G_{C},\bar{{\cal P}}%
_{\alpha}]=-i\hbar\bar{{\cal P}}_{\alpha},\label{2.3}%
\end{equation}
which corresponds to irreducible theories. BRST-invariant constraints are
defined as
\begin{equation}
{\cal T}_{\alpha}=(i\hbar)^{-1}[\Omega,\bar{{\cal P}}_{\alpha
}],\;[\Omega,{\cal T}_{\alpha}]=0.\label{2.4}%
\end{equation}

In terms of quantum antibracket (\ref{1.2}) with $\Omega$ standing for $Q$ we
have the following relations
\begin{equation}
({\cal T}_{\alpha},{\cal T}_{\beta})_{\Omega}=0,\label{2.5}%
\end{equation}
\begin{equation}
(\bar{{\cal P}}_{\alpha},\bar{{\cal P}}_{\beta})_{\Omega}=(i\hbar
)^{2}U_{\alpha\beta}^{\gamma}\bar{{\cal P}}_{\gamma}(-1)^{\varepsilon
_{\alpha}+\varepsilon_{\beta}+\varepsilon_{\gamma}},\label{2.6}%
\end{equation}
\begin{equation}
(\bar{{\cal P}}_{\alpha},{\cal T}_{\beta})_{\Omega}=\frac{1}{2}%
i\hbar\lbrack{\cal T}_{\alpha},{\cal T}_{\beta}](-1)^{\varepsilon
_{\alpha}},\label{2.7}%
\end{equation}
where structure coefficient operators
\begin{equation}
U_{\alpha\beta}^{\gamma}=-U_{\beta\alpha}^{\gamma}(-1)^{\varepsilon_{\alpha
}\varepsilon_{\beta}}\label{2.8}%
\end{equation}
are, in general, ghost-dependent.

If, in accordance with the ghost number prescriptions (\ref{2.1}),
(\ref{2.3}), we represent the operator $\Omega$ explicitly in the form of a
$C\bar{{\cal P}}$-ordered power series expansion in ghosts,
\begin{equation}
\Omega=C^{\alpha}T_{\alpha}+\sum_{n\ge1}\frac{1}{n!(n+1)!}C^{\alpha_{n+1}}\cdots
C^{\alpha_{1}}\Omega_{\alpha_{1}\cdots\alpha_{n+1}}^{\beta_{n}\cdots\beta_{1}%
}\bar{{\cal P}}_{\beta_{1}}\cdots\bar{{\cal P}}_{\beta_{n}},\label{2.9}%
\end{equation}
then the corresponding expansions for ${\cal T}_{\alpha}$ and
$U_{\alpha\beta}^{\gamma}$ are
\begin{equation}
{\cal T}_{\alpha}=T_{\alpha}+\sum_{n\ge1}\frac{1}{(n!)^{2}}C^{\alpha_{n}%
}\cdots C^{\alpha_{1}}\Omega_{\alpha_{1}\cdots\alpha_{n}\alpha}^{\beta
_{n}\cdots\beta_{1}}\bar{{\cal P}}_{\beta_{1}}\cdots\bar{{\cal P}%
}_{\beta_{n}},\label{2.10}%
\end{equation}%
\begin{equation}
U_{\alpha\beta}^{\gamma}=\sum_{n\ge0}\frac{1}{n!(n+1)!}C^{\alpha_{n}}\cdots
C^{\alpha_{1}}\Omega_{\alpha_{1}\cdots\alpha_{n}\alpha\beta}^{\gamma\beta
_{n}\cdots\beta_{1}}\bar{{\cal P}}_{\beta_{1}}\cdots\bar{{\cal P}%
}_{\beta_{n}}(-1)^{\varepsilon_{\beta}+\varepsilon_{\gamma}}.\label{2.11}%
\end{equation}
Now, due to the property (\ref{1.11}) and definition (\ref{2.4}), we get the
following commutator algebra
\begin{equation}
\lbrack{\cal T}_{\alpha},{\cal T}_{\beta}]=i\hbar U_{\alpha\beta
}^{\gamma}{\cal T}_{\gamma}-[U_{\alpha\beta}^{\gamma},\Omega]\bar{\cal
{P}}_{\gamma},\label{2.12}%
\end{equation}%
\begin{equation}
\lbrack\bar{{\cal P}}_{\alpha},\bar{{\cal P}}_{\beta}]=0,\label{2.13}%
\end{equation}%
\begin{equation}
\lbrack\bar{{\cal P}}_{\alpha},{\cal T}_{\beta}]=(i\hbar)^{-1}
(\bar{{\cal P}}_{\alpha},\bar{{\cal P}}_{\beta})_{\Omega}.\label{2.14}
\end{equation}
Thus we conclude that the BRST-invariant constraints ${\cal T}_{\alpha}$
together with ghost momenta $\bar{{\cal P}}_{\alpha}$ form two dual
operator algebras, namely, the quantum-antibracket algebra (\ref{2.5}) -
(\ref{2.7}) and commutator algebra (\ref{2.12}) - (\ref{2.14}).

\section{Generating equations of BRST-invariant constraint algebra.}

As we have established that ${\cal T}_{\alpha}$ together with
$\bar{{\cal P}}_{\alpha}$ form two dual algebras, it seems quite natural to
formulate the corresponding generating equations, in the line of general
ideology of BRST-BFV approach. We can regard ${\cal T}_{\alpha}$ and
$\bar{{\cal P}}_{\alpha}$ as first-class constraints with (\ref{2.12}) -
(\ref{2.14}) being their involution relations. Moreover, we can rotate these
first-class constraints with some (nonsingular) matrices, so that it seems
natural to generalize a little bit the definition of ${\cal T}_{\alpha}.$

First of all, let us rotate $\bar{{\cal P}}_{\alpha}$ in (\ref{2.4}),
\begin{equation}
\bar{{\cal P}}_{\alpha}\;\rightarrow\;X_{\alpha}=\Lambda_{\alpha}^{\beta
}\bar{{\cal P}}_{\beta},\label{3.1}%
\end{equation}
so that new ${\cal T}_{\alpha}$ read
\begin{equation}
{\cal T}_{\alpha}=(i\hbar)^{-1}[\Omega,X_{\alpha}],\quad\;[G_{C},X_{\alpha
}]=-i\hbar X_{\alpha}.\label{3.2}%
\end{equation}
These ${\cal T}_{\alpha}$, however, remain strongly BRST-invariant,
$[\Omega,{\cal T}_{\alpha}]=0$. To weaken the invariance, we can modify the
definition of ${\cal T}_{\alpha}$ yet more,
\begin{equation}
{\cal T}_{\alpha}=(i\hbar)^{-1}[\Omega,X_{\alpha}]-V_{\alpha}^{\beta
}X_{\beta}(-1)^{\varepsilon_{\alpha}+\varepsilon_{\beta}},\label{3.3}%
\end{equation}
with $V_{\alpha}^{\beta}$ being a flat BRST connection,
\begin{equation}
R_{\alpha}^{\beta}\equiv(i\hbar)^{-1}[\Omega,V_{\alpha}^{\beta}]-
V_{\alpha}^{\gamma}V_{\gamma}^\beta
(-1)^{\varepsilon_{\alpha}+\varepsilon_{\gamma}}=0.\label{3.4}%
\end{equation}
Then we have a weak BRST invariance,
\begin{equation}
\lbrack\Omega,{\cal T}_{\alpha}]=i\hbar V_{\alpha}^{\beta}{\cal T}%
_{\beta},\label{3.5}%
\end{equation}
which corresponds to the rotation
\begin{equation}
{\cal T}_{\alpha}\rightarrow G_{\alpha}^{\beta}{\cal T}_{\beta
},\;X_{\alpha}\rightarrow G_{\alpha}^{\beta}X_{\beta}(-1)^{\varepsilon
_{\alpha}+\varepsilon_{\beta}},\label{3.6}%
\end{equation}
in (\ref{3.2}), together with the choice
\begin{equation}
V_{\alpha}^{\beta}=(i\hbar)^{-1}[\Omega,G_{\alpha}^{-1\gamma}]G_{\gamma
}^{\beta}.\label{3.7}%
\end{equation}
We expect the above rotations (\ref{3.1}) - (\ref{3.6}) to be a part of
natural arbitrariness in the general solution to the algebra-generating
equations.

Now, let us turn directly to the formulation of generating equations in
question. We begin with some operators ${\cal T}_{\alpha}$ and $X_\alpha$
living in the same extended phase space as a BRST operator $\Omega$ does.
Their Grassmann parities are
\begin{equation}
\varepsilon({\cal T}_{\alpha})=\varepsilon_{\alpha},\;\varepsilon
(X_{\alpha})=\varepsilon_{\alpha}+1,\label{3.8}%
\end{equation}
and their intrinsic ghost number values are given by
\begin{equation}
\lbrack G_{C},{\cal T}_{\alpha}]=0,\;[G_{C},X_{\alpha}]=-i\hbar X_{\alpha
}.\label{3.9}%
\end{equation}
Next, let us extend the phase space yet more by introducing new ghost-type
canonical pairs via the correspondence
\begin{equation}
{\cal T}_{\alpha}\mapsto(B^{\alpha},\Pi_{\alpha}),\;\;X_{\alpha}
\mapsto(B_{\alpha}^{\ast},\Pi_{\ast}^{\alpha}),\label{3.10}%
\end{equation}
with the only nonzero commutators
\begin{equation}
\lbrack B^{\alpha},\Pi_{\beta}]=i\hbar\delta_{\beta}^{\alpha},\;\;[B_{\alpha
}^{\ast},\Pi_{\ast}^{\beta}]=i\hbar\delta_{\alpha}^{\beta}.\label{3.11}%
\end{equation}
Their Grassmann parities are
\begin{equation}
\varepsilon(B^{\alpha})=\varepsilon(\Pi_{\alpha})=\varepsilon_{\alpha
}+1,\;\varepsilon(B_{\alpha}^{\ast})=\varepsilon(\Pi_{\ast}^{\alpha
})=\varepsilon_{\alpha}.\label{3.12}%
\end{equation}
All new operators commute with the intrinsic ghost number operator $G_{C}$.
However, they have their own ghost number operators $G_{B}$ and
$G_{B^{\ast}},$%
\begin{equation}
\lbrack G_{B},B^{\alpha}]=i\hbar B^{\alpha},\;\;[G_{B},\Pi_{\alpha}%
]=-i\hbar\Pi_{\alpha},\;\;\label{3.13}%
\end{equation}%
\begin{equation}
\lbrack G_{B^{\ast}},B_{\alpha}^{\ast}]=i\hbar B_{\alpha}^{\ast}%
,\;\;[G_{B^{\ast}},\Pi_{\ast}^{\alpha}]=-i\hbar\Pi_{\ast}^{\alpha
},\label{3.14}%
\end{equation}%
\begin{equation}
[G_{B^{\ast}},B^{\alpha}]=[G_{B^{\ast}},\Pi_{\alpha}]=[G_{B},B_{\alpha
}^{\ast}]=[G_{B},\Pi_{\ast}^{\alpha}]=0, \label{3.15}%
\end{equation}
\begin{equation}
\lbrack G_{B},G_{B^*}]=[G_{C},G_{B}]=[G_{C},G_{B^{\ast}}]=0.
\label{3.15a}%
\end{equation}
A total ghost number operator is
\begin{equation}
G=G_{C}+G_{B}-2G_{B^{\ast}}=G_{CB^{\ast}}+G_{BB^{\ast}},\label{3.16}%
\end{equation}
where
\begin{equation}
G_{CB^{\ast}}=G_{C}-G_{B^{\ast}},\;\;G_{BB^{\ast}}=G_{B}-G_{B^\ast}.
\label{3.17}%
\end{equation}

Let $A$ be an arbitrary operator. We define the total ghost number value,
gh($A$), and total degree, deg($A$), as
\begin{equation}
\lbrack
G,A]=i\hbar\,\hbox{gh}(A)\,A,\;\;[G_{BB^{\ast}},A]=i\hbar\,
\hbox{deg}(A)\,A,\label{3.18}%
\end{equation}
so that
\begin{equation}
\lbrack G_{CB^{\ast}},A]=i\hbar\left(\hbox{gh}(A)-\hbox{deg}(A)\right)A.
\label{3.19}%
\end{equation}
We have, in particular,
\begin{equation}
\varepsilon(\Omega)=1,\;\hbox{gh}(\Omega)=1,\;\hbox{deg}
(\Omega)=0.\label{3.20}%
\end{equation}

In what follows, it is convenient to use the condensed notation
\begin{equation}
T_{A}\equiv\left\{{\cal T}_{\alpha};-X_{\alpha}\right\},
\label{3.30}%
\end{equation}
\begin{equation}
C^{A}\equiv\left\{  B^{\alpha};\Pi_{\ast}^{\alpha}
(-1)^{\varepsilon_{\alpha}+1}\right\},\quad
\bar{{\cal P}}_{A}\equiv\left\{\Pi_{\alpha};B_{\alpha}^{\ast}\right\},\quad
[C^{A},\bar{{\cal P}}_{B}]=i\hbar\delta_{B}^{A}.\label{3.34}
\end{equation}
We have
\begin{equation}\label{3.31}
\varepsilon(T_{A})=\left\{\varepsilon_{\alpha};
\varepsilon_{\alpha }+1\right\},\quad
\hbox{gh}(T_{A})=\left\{0;-1\right\},\quad
\hbox{deg}(T_{A})=\left\{0;0\right\},
\end{equation}
\begin{equation}
\varepsilon(C^{A})=\left\{\varepsilon_{\alpha} +1;
\varepsilon_{\alpha}\right\},\quad
\hbox{gh}(C^{A})=\left\{1;2\right\},\quad
\hbox{deg}(C^{A})=\left\{1;1\right\},\label{3.32}
\end{equation}
\begin{equation}\label{3.32a}
\varepsilon(T_{A})\equiv\varepsilon_A,\quad
\varepsilon(\bar{\cal P}_{A})=\varepsilon(C^A)=\varepsilon_A+1,
\end{equation}
\begin{equation}
\hbox{gh}(\bar{\cal P}_{A})=-\hbox{gh}(C^{A}),\quad
\hbox{deg}(\bar{\cal P}_{A})=-\hbox{deg}(C^{A}).\label{3.32b}
\end{equation}

In the new extended phase space, spanned by original phase variables,
ordinary ghosts and new variables (\ref{3.10}), let us consider the following
set of equations
\begin{equation}
\lbrack\Sigma_{1},\Sigma_{1}]=0,\;[\Delta,\Delta]=0,\;[\Delta,\Sigma
_{1}]=0,\label{3.21}%
\end{equation}%
\begin{equation}
\varepsilon(\Sigma_{1})=1,\;\hbox{gh}(\Sigma_{1})=1,\;\hbox{deg}
(\Sigma_{1})=1,\label{3.22}%
\end{equation}%
\begin{equation}
\varepsilon(\Delta)=1,\;\hbox{gh}(\Delta)=1,\;\hbox{deg}
(\Delta)=0,\label{3.23}%
\end{equation}
together with the boundary conditions
\begin{equation}\label{3.24}
\Sigma_{1}=C^AT_A+\ldots,\quad
\Delta=\Omega+\ldots,
\end{equation}
where dots, $\ldots$, mean all possible higher-order terms in
$(C^A,\bar{\cal P}_A)$, allowed by (\ref{3.22}), (\ref{3.23}).

We also require for the operator $\Delta$ to satisfy the extra condition:
the $\Delta$-antibracket matrix $(B^{\alpha},B^*_{\beta})_{\Delta}$ should be
invertible.

We state that the equations (\ref{3.21}) - (\ref{3.24}), when expanded in
$(C^A,\bar{\cal P}_A)$, generate a BRST-invariant constraint algebra.

In order to see this, let  us consider the $C\bar{\cal P}$-ordered expansions
for $\Sigma_1$ and $\Delta$
\begin{equation}\label{12}
\Sigma_1=C^AT_A+\frac{1}{2}C^BC^AU^C_{AB}\bar{\cal P}_C
(-1)^{\varepsilon_B+\varepsilon_C}+\ldots,
\end{equation}
\begin{equation}\label{13}
\Delta=\Omega+C^AV_A^B\bar{\cal P}_B(-1)^{\varepsilon_B}+
\frac{1}{4}C^BC^AV^{CD}_{AB}\bar{\cal P}_D\bar{\cal P}_C
(-1)^{\varepsilon_B+\varepsilon_D}+\ldots.
\end{equation}

By substituting (\ref{12}) into the first in (\ref{3.21}), we get, in the
second order in $C^A$, the standard involution relations,
\begin{equation}\label{14}
[T_A,T_B]=i\hbar U^C_{AB}T_C.
\end{equation}
Next, by substituting (\ref{13}) into the second in (\ref{3.21}), we get,
in the zeroth and first order in $C^A$,
\begin{equation}\label{15}
[\Omega,\Omega]=0,
\end{equation}
and
\begin{equation}\label{16}
[V^B_A,\Omega](-1)^{\varepsilon_B}=i\hbar V_A^CV^B_C.
\end{equation}
In the same way, by substituting (\ref{12}), (\ref{13}) into the third in
(\ref{3.21}), we get, in the first order in $C^A$,
\begin{equation}\label{17}
[T_A,\Omega]=-i\hbar V^B_AT_B.
\end{equation}
In (\ref{15}) we recognize the nilpotence condition for $\Omega$. It is
remarkable that (\ref{16}) is nothing but the nilpotence condition for
matrix-extended $\Omega$, $\hat{\Omega}^B_A$,
\begin{equation}\label{18}
\hat{\Omega}_A^C\hat{\Omega}_C^B=0,\quad
\hat{\Omega}_A^B\equiv(-1)^{\varepsilon_A}\delta^B_A\Omega-i\hbar V^B_A.
\end{equation}

In their turn, the involution relations (\ref{17}) determine $T_A$ to be
BRST-invariant constraints in the most general (weak) form.

It is easy to see that the previous (particular) representations (\ref{3.2}),
(\ref{3.3}), (\ref{3.5}) follow immediately from (\ref{17}) when choosing
$V^B_A$ in the form
\begin{equation}\label{20}
V^B_A=\left(\begin{array}{cc}V^\beta_\alpha(-1)^{\varepsilon_\alpha}&0\\
\delta^\beta_\alpha(-1)^{\varepsilon_\alpha}&
-V^\beta_\alpha(-1)^{\varepsilon_\beta}
\end{array}\right).
\end{equation}

Now, let us consider the second in (\ref{3.21}) in the second order in $C^A$.
We get
$$
[V^C_A,V^D_B](-1)^{(\varepsilon_B+1)(\varepsilon_C+1)}-(A\leftrightarrow B)
(-1)^{\varepsilon_A\varepsilon_B}=i\hbar(V_A^EV^{CD}_{EB}(-1)^{\varepsilon_B}
-(A\leftrightarrow B)(-1)^{\varepsilon_A\varepsilon_B})+
$$
\begin{equation}\label{21}
+i\hbar(V^{CE}_{AB}V^D_E(-1)^{\varepsilon_C}
-(C\leftrightarrow D)(-1)^{\varepsilon_C\varepsilon_D})-
[V^{CD}_{AB},\Omega](-1)^{\varepsilon_C+\varepsilon_D}-
\frac{1}{2}(i\hbar)^2V^{EF}_{AB}V^{CD}_{FE}.
\end{equation}

In the same order in $C^A$, the third in (\ref{3.21}) yields
$$
([T_A,V^C_B]-(A\leftrightarrow B)(-1)^{\varepsilon_A\varepsilon_B})-
i\hbar U^D_{AB}V^C_D+
$$
\begin{equation}\label{22}
+i\hbar(V^D_AU^C_{DB}(-1)^{\varepsilon_B}
-(A\leftrightarrow B)(-1)^{\varepsilon_A\varepsilon_B})+
[U^C_{AB},\Omega](-1)^{\varepsilon_C}+\frac{1}{2}i\hbar
V^{ED}_{AB}Z^C_{DE}=0,
\end{equation}
where
\begin{equation}\label{23}
Z^C_{AB}\equiv T_A\delta^C_B-T_B\delta^C_A(-1)^{\varepsilon_A\varepsilon_B}-
i\hbar U^C_{AB},
\end{equation}
\begin{equation}\label{24}
Z^C_{AB}T_C=0.
\end{equation}
By multiplying (\ref{22}) by $T_C$ from the right, we get, identically, zero
due to (\ref{14}), (\ref{17}).

There are no more equations up to the third order in $C^A$.

Given first-class constraints $T_A$, eqs. (\ref{14}) determine $U^C_{AB}$.
Then, eqs. (\ref{16}), (\ref{17}) determine $V^A_B$. Then, eqs. (\ref{21}),
(\ref{22}) determine $V^{CD}_{AB}$, and so on.

If generating equations (\ref{3.21}) allow for $\Delta$ linear in $C^A$,
\begin{equation}\label{25}
\Delta=\Omega+C^AV_A^B\bar{\cal P}_B(-1)^{\varepsilon_B},
\end{equation}
which implies, in accordance with (\ref{21}), (\ref{22}), that
\begin{equation}\label{26}
[V^C_A,V^D_B](-1)^{(\varepsilon_B+1)(\varepsilon_C+1)}-(A\leftrightarrow B)
(-1)^{\varepsilon_A\varepsilon_B}=0,
\end{equation}
and
$$
([T_A,V^C_B]-(A\leftrightarrow B)(-1)^{\varepsilon_A\varepsilon_B})-
i\hbar U^D_{AB}V^C_D+
$$
\begin{equation}\label{27}
+i\hbar(V^D_AU^C_{DB}(-1)^{\varepsilon_B}
-(A\leftrightarrow B)(-1)^{\varepsilon_A\varepsilon_B})+
[U^C_{AB},\Omega](-1)^{\varepsilon_C}=0,
\end{equation}
then, for any quantities $f$, $g$, $h$, depending on $B^\alpha$,
$B^*_\alpha$ only, their $\Delta$-antibrackets satisfy
\begin{equation}\label{28}
[(f,g)_\Delta,h]=0,
\end{equation}
so that their 3-antibrackets vanish
\begin{equation}\label{29}
(f,g,h)_\Delta=0,
\end{equation}
and, thereby, Jacobi identities (\ref{1.6}) become strong.
As these $f$, $g$, $h$ commute among themselves, Leibniz rule (\ref{1.5})
becomes strong as well. Besides, we have
\begin{equation}
\varepsilon(B^{\alpha})+\varepsilon(B_{\alpha}^{\ast})=1,\label{3.27}
\end{equation}
\begin{equation}
\hbox{gh}(B^{\alpha})+\hbox{gh}(B_{\alpha}^{\ast})=-1,\label{3.28}
\end{equation}
\begin{equation}
\hbox{deg}(B^{\alpha})+\hbox{deg}(B_{\alpha}^{\ast})=0,\label{3.29}
\end{equation}
by assignment.
So, it follows from (\ref{29}) - (\ref{3.29}) that the variables $B^\alpha$
and $B^*_\alpha$ behave as normal fields and antifields with respect to
$\Delta$-antibracket, provided the conditions (\ref{26}), (\ref{27}) are
satisfied.

We emphasize, however, that the conditions (\ref{26}), (\ref{27}) are not
required imperatively to be fulfilled in any case. They merely specify a
certain basis of constraints $T_A$ and quantities $V^B_A$, in which the
formalism allows for a simple interpretation to the variables $B^\alpha$,
$B^*_\alpha$. In the general case, the coefficients $V^{CD}_{AB}$ are
nonzero, and the expansions (\ref{12}), (\ref{13}) involve all higher orders
in ghosts. Therefore, $\Delta$-antibrackets do not meet, in general, a strong
Jacobi identity, even if their entries depend on $B^\alpha$, $B^*_\alpha$
only.

In principle, the involution relations (\ref{14}) - (\ref{17}) are the only
conditions the lowest-order terms in (\ref{12}), (\ref{13}) should satisfy
to. However, we require for r.h.s. in (\ref{17}) to resolve for ${\cal
T}_\alpha$: this is just the extra condition formulated below (\ref{3.24}).
This condition means that any constraints $T_A$, satisfying these involution
relations, can be rotated with a nonsingular matrix to take the form
$T_A=\{(i\hbar)^{-1}[\Omega,\bar{\cal P}_\alpha];-\bar{\cal P}_\alpha\}$.

\section{Generating equations of antibracket algebra}

As we have seen above, the variables $B^\alpha$ and $B^*_\alpha$ behave as
fields and antifields with respect to $\Delta$-antibracket. It seems quite
natural to expect a similar behaviour for momenta $\Pi^\alpha_*$ and
$\Pi_\alpha$ with respect to some ``dual'' antibracket.

To put the above idea into effect, let us define the resolvent operator
$\bar{\Delta}$ to satisfy the generating equations
\begin{equation}\label{4.1}
[\bar{\Delta},\bar{\Delta}]=0,\quad [\Delta,\bar{\Delta}]=i\hbar G_{BB^*},
\end{equation}

\begin{equation}\label{4.2}
\varepsilon(\bar{\Delta})=1,\quad \hbox{gh}(\bar{\Delta})=-1,\quad
\hbox{deg}(\bar{\Delta})=0,
\end{equation}
together with the boundary condition
\begin{equation}\label{4.3}
\bar{\Delta}=\bar{\Omega}+\ldots,
\end{equation}
where dots, $\ldots$, mean  all possible higher order terms in the variables
$(C^A,\bar{\cal P}_A)$, allowed by (\ref{4.2}), while $\bar{\Omega}$ is of
the zeroth order.

Let us consider for $\bar{\Delta}$ the $C\bar{\cal P}$-ordered power series
expansion
\begin{equation}\label{4.4}
\bar{\Delta}=\bar{\Omega}+C^A\bar{V}_A^B\bar{\cal P}_B(-1)^{\varepsilon_B}+
\frac{1}{4}C^BC^A\bar{V}^{CD}_{AB}\bar{\cal P}_D\bar{\cal P}_C
(-1)^{\varepsilon_B+\varepsilon_D}+\ldots,
\end{equation}
similar to (\ref{13}). Then, we get from (\ref{4.1}), (\ref{4.2}) the
following lowest-order equations for coefficient operators
\begin{equation}\label{4.5}
[\bar{\Omega},\bar{\Omega}]=0,
\end{equation}
\begin{equation}\label{4.6}
[\bar{V}^B_A,\bar{\Omega}](-1)^{\varepsilon_B}=i\hbar\bar{V}^C_A\bar{V}^B_C,
\end{equation}
\begin{equation}\label{4.7}
[\Omega,\bar{\Omega}]=0,
\end{equation}
\begin{equation}\label{4.8}
[V^B_A,\bar{\Omega}](-1)^{\varepsilon_B}+
[\bar{V}^B_A,\Omega](-1)^{\varepsilon_B}+i\hbar\delta^B_A=
i\hbar V^C_A\bar{V}^B_C+i\hbar\bar{V}^C_AV^B_C.
\end{equation}
In terms of the operator-valued matrix (\ref{18}) and the same for
$\bar{\Delta}$,
\begin{equation}\label{4.9}
\hat{\bar{\Omega}}_A^B\equiv(-1)^{\varepsilon_A}\delta^B_A\bar{\Omega}-
i\hbar\bar{V}^B_A,
\end{equation}
equations (\ref{4.5}) - (\ref{4.8}) rewrite as
\begin{equation}\label{4.10}
\hat{\bar{\Omega}}_A^C\hat{\bar{\Omega}}_C^B=0,
\end{equation}
\begin{equation}\label{4.11}
\hat{\Omega}_A^C\hat{\bar{\Omega}}_C^B+
\hat{\bar{\Omega}}_A^C\hat{\Omega}_C^B=i\hbar\delta^B_A.
\end{equation}
As for the nilpotent operator $\bar{\Omega}$, it lives in the same phase
space as $\Omega$ does, and, when expanded in ordinary ghosts $(C^\alpha,
\bar{\cal P}_\alpha)$, begins with $\bar{\Omega}=\bar{T}^\alpha
\bar{\cal P}_\alpha(-1)^{\varepsilon_\alpha}+\ldots$, where $\bar{T}^\alpha$
are linear combinations of the first-class constraints $T_\alpha$, dual to
$T_\alpha$, $\bar{T}^\alpha T_\alpha=0$.

The same as for $\Delta$, if generating equations allow for $\bar{\Delta}$
linear in $C^A$, then, for any quantities depending on $\Pi^\alpha_*$,
$\Pi_\alpha$ only, $\bar{\Delta}$-antibracket meets a strong Jacobi identity.

However, now we have, by assignment, a counterpart of (\ref{3.27}) -
(\ref{3.29}) in the form
\begin{equation}
\varepsilon(\Pi_*^{\alpha})+\varepsilon(\Pi_{\alpha})=1,\label{4.12}
\end{equation}
\begin{equation}
\hbox{gh}(\Pi_*^{\alpha})+\hbox{gh}(\Pi_{\alpha})=1,\label{4.13}
\end{equation}%
\begin{equation}
\hbox{deg}(\Pi_*^{\alpha})+\hbox{deg}(\Pi_{\alpha})=0,\label{4.14}
\end{equation}
We see that the signs in r.h.s. of (\ref{3.28}) and (\ref{4.13}) are opposite,
which means that, in contrast to $B^\alpha$, $B^*_\alpha$, the momenta
$\Pi^\alpha_*$ and $\Pi_\alpha$ behave as ``twisted'' fields and antifields
\cite{CatFel00} - \cite{BatMar02}.

In what follows, we imply that a solution to the generating equations
(\ref{3.21}) - (\ref{3.24}) and (\ref{4.1}) - (\ref{4.3}) does exist.

Then, by commuting $\bar{\Delta}$ with the third equation in (\ref{3.21}),
and using the third in (\ref{3.22}) and the second in (\ref{4.1}), we get

\begin{equation}\label{4.8a}
\Sigma_1=(i\hbar)^{-1}[\Delta,S_1],
\end{equation}
where
\begin{equation}\label{4.9a}
S_1=(i\hbar)^{-1}[\bar{\Delta},\Sigma_1]+(i\hbar)^{-1}[\Delta,Y_1],
\end{equation}
and $Y_1$ is an arbitrary Fermionic operator with gh$(Y_1)=-1$, deg$(Y_1)=1$.

For $S_1$ itself, we have
\begin{equation}\label{4.10a}
\varepsilon(S_1)=0,\quad \hbox{gh}(S_1)=0,\quad \hbox{deg}(S_1)=1.
\end{equation}

By substituting (\ref{4.8a}) into the first in (\ref{3.21}), and using the
property (\ref{1.11}), we get
\begin{equation}\label{4.11a}
[\Delta,(S_1,S_1)_\Delta]=0.
\end{equation}
In its turn, by commuting $\bar{\Delta}$ with (\ref{4.11a}), we obtain,
similarly to (\ref{4.8a}),
\begin{equation}\label{4.12a}
(S_1,S_1)_\Delta=i\hbar[\Delta,S_2],
\end{equation}
where
\begin{equation}\label{4.13a}
S_2=\frac{1}{2}(i\hbar)^{-3}[\bar{\Delta},(S_1,S_1)_\Delta]+
(i\hbar)^{-1}[\Delta,Y_2],
\end{equation}
and $Y_2$ is an arbitrary Fermionic operator with gh$(Y_2)=-1$, deg$(Y_2)=2$.

For $S_2$ itself, we have
\begin{equation}\label{4.14a}
\varepsilon(S_2)=0,\quad \hbox{gh}(S_2)=0,\quad \hbox{deg}(S_2)=2.
\end{equation}

Now, let us consider the following master equation
\begin{equation}\label{4.15}
(S,S)_\Delta=i\hbar[\Delta,S],
\end{equation}
for a Bosonic operator $S$ of the form
\begin{equation}\label{4.16}
S=\sum_{k\ge0}S_k,\quad S_0=G_{CB^*},\quad \varepsilon(S_k)=0,\quad
\hbox{gh}(S_k)=0,\quad \hbox{deg}(S_k)=k.
\end{equation}
We have
\begin{equation}\label{4.17}
[S_0,\Delta]=i\hbar\Delta,\quad [S_0,S_k]=-i\hbar kS_k.
\end{equation}
By substituting (\ref{4.16}) into (\ref{4.15}), and using (\ref{4.17}), we
get the following chain of equations
\begin{equation}\label{4.18}
F_k=0,\quad k\ge2,
\end{equation}
where
\begin{equation}\label{4.19}
F_k\equiv R_k-i\hbar(k-1)[\Delta,S_k],
\end{equation}
\begin{equation}\label{4.20}
R_k\equiv\sum_{j=1}^{k-1}(S_j,S_{k-j})_\Delta.
\end{equation}
At $k=2$, (\ref{4.18}) coincides exactly with (\ref{4.12a}), so that we can
identify $S_1$ and $S_2$ in (\ref{4.16}) with (\ref{4.9a}) and (\ref{4.13a}),
respectively. Then, by making use of the identity (\ref{1.10}) for
$Q=\Delta$, $B=S$, it is easy to show that
\begin{equation}\label{4.21}
[\Delta,R_k]=0,
\end{equation}
provided the equations
\begin{equation}\label{4.22}
F_m=0,\quad m=2,\ldots,k-1,
\end{equation}
are satisfied.

Indeed, it follows from (\ref{1.10}) that
\begin{equation}\label{4.23}
6(S,F)_\Delta-[[S,F],\Delta]=4i\hbar[\Delta,F],
\end{equation}
where
\begin{equation}\label{4.24}
F\equiv(S,S)_\Delta-i\hbar[\Delta,S].
\end{equation}
Let the equations (\ref{4.22}) be satisfied. Then, by taking in (\ref{4.23})
the sector with degree equal to $k$, we get
\begin{equation}\label{4.25}
6(S_0,F_k)_\Delta-[[S_0,F_k],\Delta]=4i\hbar[\Delta,F_k],
\end{equation}
which yields immediately
\begin{equation}\label{4.26}
(k-2)[\Delta,R_k]=0.
\end{equation}
Finally, by commuting $\bar{\Delta}$ with (\ref{4.21}), we obtain
(\ref{4.18}) with
\begin{equation}\label{4.27}
S_k=\frac{1}{k(k-1)}(i\hbar)^{-3}[\bar{\Delta},R_k]+(i\hbar)^{-1}[\Delta,Y_k],
\end{equation}
\begin{equation}\label{4.28}
\varepsilon(Y_k)=1,\quad \hbox{gh}(Y_k)=-1,\quad \hbox{deg}(Y_k)=k.
\end{equation}
Thus, we conclude that all the operators $S_k$ entering the expansion
(\ref{4.16}) for $S$ do exist. Thereby, we have established that master
equation (\ref{4.15}) has a solution generated by $\Sigma_1$ via (\ref{4.8a}),
(\ref{4.12a}). This solution describes the antibracket algebra generated by
BRST-invariant constraints.

Let us consider the simplest case of a rank-one theory, a Lie-type algebra
with constant structure coefficients. Then, by choosing
$C\bar{\cal P}$-ordering in ghost sector, we have the following BRST operator
$\Omega$,
\begin{equation}\label{4.29}
\Omega=C^\alpha
T_\alpha+\frac{1}{2}C^\beta C^\alpha U^\gamma_{\alpha\beta} \bar{\cal
P}_\gamma(-1)^{\varepsilon_\beta+\varepsilon_\gamma},
\end{equation}
with $U^\gamma_{\alpha\beta}$ being constant.

Consider the simplest possible form of the operator $\Delta$, which is
\begin{equation}\label{4.30}
\Delta=\Omega+\Pi^\alpha_*\Pi_\alpha(-1)^{\varepsilon_\alpha+1},
\end{equation}
so that the corresponding resolvent operator $\bar{\Delta}$ reads
\begin{equation}\label{4.31}
\bar{\Delta}=\bar{\Omega}-B^\alpha B^*_\alpha.
\end{equation}
With an operator $\Delta$ chosen in the form (\ref{4.30}), the equation
(\ref{4.12a}) has a solution of the form
\begin{equation}\label{4.32}
S_1=\bar{\cal P}_\alpha B^\alpha+\frac{1}{2}B^\beta
B^\alpha U^\gamma_{\alpha\beta}
B^*_\gamma(-1)^{\varepsilon_\beta},
\end{equation}
\begin{equation}\label{4.33}
(S_1,S_1)_\Delta=0,\quad S_2=0,
\end{equation}
while the nilpotent operator $\Sigma_1$ in (\ref{4.8a}) is given by the formula
$$
\Sigma_1=(i\hbar)^{-1}[\Delta,S_1]=
$$
\begin{equation}
=B^\alpha{\cal T}_\alpha+\frac{1}{2}B^\beta B^\alpha U^\gamma_{\alpha\beta}
\Pi_\gamma(-1)^{\varepsilon_\beta+\varepsilon_\gamma}+
\Pi^\alpha_*\bar{\cal P}_\alpha(-1)^{\varepsilon_\alpha}-
\Pi_*^\beta B^\alpha U^\gamma_{\alpha\beta}B^*_\gamma(-1)^{\varepsilon_\beta},
\label{4.34}
\end{equation}
where ${\cal T}_\alpha$ are $C\bar{\cal P}$-ordered BRST-invariant
constraints,
\begin{equation}\label{4.35}
{\cal T}_\alpha=(i\hbar)^{-1}[\Omega,\bar{\cal P}_\alpha]=T_\alpha+
C^\beta U^\gamma_{\alpha\beta}\bar{\cal P}_\gamma
(-1)^{\varepsilon_\alpha+\varepsilon_\gamma}.
\end{equation}

In the general case, it can be shown that the appearance of nonzero $S_k$,
$k\ge2$, entering the expansion (\ref{4.16}), is an effect of anticanonical
transformation (\ref{1.14}) - (\ref{1.16}) applied to the operator $S_1$
satisfying the homogeneous master equation (\ref{4.33}). Roughly speaking, we
can say that r.h.s. of (\ref{4.12a}) comes just from the deviation
(\ref{1.19}).

Let us also mention that the solution (\ref{4.32}), (\ref{4.33}) remains valid
even if structure coefficients $U^\gamma_{\alpha\beta}$ in (\ref{4.29}) are
not constant but satisfy the quasigroup conditions
\begin{equation}\label{4.36}
[U^\gamma_{\alpha\beta},U^\rho_{\mu\nu}]=0,\quad
[[T_\alpha,U^\delta_{\beta\gamma}],U^\rho_{\mu\nu}]=0.
\end{equation}

\section{BRST-invariant constraint algebra in rank-one theories}

Here, we give some explicit formulas potentially useful for practical
applications to rank-one theories. We consider BRST-invariant algebra in its
commutator and antibracket form in the cases of Weyl- and Wick- ordered
ghost sector, which are most popular ones.

\subsection{Weyl-ordered ghost sector}

In the case of Weyl-ordered ghost sector, a rank-one theory is described
by the following BRST-operator linear in ghost momenta \cite{BatFra88},
$$
\Omega=C^\alpha T_\alpha+
\frac{1}{6}C^\beta C^\alpha U^\gamma_{\alpha\beta}\bar{\cal P}_\gamma
(-1)^{\varepsilon_\beta+\varepsilon_\gamma}+
$$
\begin{equation}
+\frac{1}{6}C^\alpha U^\gamma_{\alpha\beta}\bar{\cal P}_\gamma C^\beta
(-1)^{\varepsilon_\beta+\varepsilon_\gamma}+
\frac{1}{6}\bar{\cal P}_\gamma U^\gamma_{\alpha\beta} C^\beta C^\alpha
(-1)^{\varepsilon_\beta+\varepsilon_\gamma+
(\varepsilon_\alpha+\varepsilon_\beta)(\varepsilon_\gamma+1)}.
\label{5.1}
\end{equation}
 Original constraint algebra is given by the involution relation
 \cite{BatFra88},
\begin{equation}\label{5.2}
[T_\alpha,T_\beta]=\frac{i\hbar}{2}\left(U^\gamma_{\alpha\beta}T_\gamma+
T_\gamma U^\gamma_{\alpha\beta}
(-1)^{(\varepsilon_\alpha+\varepsilon_\beta+1)\varepsilon_\gamma}\right)+
\left(\frac{i\hbar}{2}\right)^2
[U^\gamma_{\alpha\delta},U^\delta_{\gamma\beta}]
(-1)^{\varepsilon_\delta(\varepsilon_\beta+1)}.
\end{equation}

BRST-invariant constraints are
\begin{equation}\label{5.3}
{\cal T}_\alpha=(i\hbar)^{-1}[\Omega,\bar{\cal P}_\alpha]=T_\alpha+
\frac{1}{2}\left(C^\beta U^\gamma_{\beta\alpha}\bar{\cal P}_\gamma
(-1)^{\varepsilon_\alpha+\varepsilon_\gamma}+
\bar{\cal P}_\gamma U^\gamma_{\alpha\beta}C^\beta
(-1)^{\varepsilon_\alpha+
(\varepsilon_\alpha+\varepsilon_\beta+1)\varepsilon_\gamma}\right).
\end{equation}
Their commutator algebra reads
$$
[{\cal T}_\alpha,{\cal T}_\beta]=\frac{i\hbar}{2}\left(
U^\gamma_{\alpha\beta}{\cal T}_\gamma+ {\cal T}_\gamma U^\gamma_{\alpha\beta}
(-1)^{(\varepsilon_\alpha+\varepsilon_\beta+1)\varepsilon_\gamma}\right)+
$$
\begin{equation}
+\frac{1}{2}\left([\Omega,U^\gamma_{\alpha\beta}]\bar{\cal P}_\gamma
(-1)^{\varepsilon_\alpha+\varepsilon_\beta+\varepsilon_\gamma}-
\bar{\cal P}_\gamma [\Omega,U^\gamma_{\alpha\beta}]
(-1)^{(\varepsilon_\alpha+\varepsilon_\beta)\varepsilon_\gamma}\right),
\label{5.4}
\end{equation}
\begin{equation}\label{5.5}
[\bar{\cal P}_\alpha,\bar{\cal P}_\beta]=0,\quad
[\bar{\cal P}_\alpha,{\cal T}_\beta]=i\hbar U^\gamma_{\alpha\beta}\bar{\cal
P}_\gamma (-1)^{\varepsilon_\alpha+\varepsilon_\beta+\varepsilon_\gamma}.
\end{equation}
We see that the extension, represented by the third term in r.h.s. in
(\ref{5.2}), is absent in (\ref{5.4}), although we have in (\ref{5.4}) an
admixture of ghost momenta $\bar{\cal P}_\alpha$, instead.

The antibracket algebra, corresponding to (\ref{5.4}), (\ref{5.5}), reads
\begin{equation}\label{5.6}
({\cal T}_\alpha,{\cal T}_\beta)_\Omega=0,\quad
(\bar{\cal P}_\alpha,\bar{\cal P}_\beta)_\Omega=(i\hbar)^2
U^\gamma_{\alpha\beta}\bar{\cal P}_\gamma
(-1)^{\varepsilon_\alpha+\varepsilon_\beta+\varepsilon_\gamma},
\end{equation}
\begin{equation}\label{5.7}
(\bar{\cal P}_\alpha,{\cal T}_\beta)_\Omega=\frac{1}{2}i\hbar
[{\cal T}_\alpha,{\cal T}_\beta](-1)^{\varepsilon_\alpha}.
\end{equation}

\subsection{Wick-ordered ghost sector}

As usual, Wick ghost sector is represented by two sets of Wick pairs,
($C^\alpha,\bar{C}^\dagger_\alpha$) and
($\bar{C}_\alpha,C^{\dagger\alpha}$), with the only nonzero
commutators,
\begin{equation}\label{5.8}
[C^\alpha,\bar{C}^\dagger_\beta]=\delta^\alpha_\beta,\quad
[\bar{C}_\alpha,C^{\dagger\beta}]=\delta_\alpha^\beta.
\end{equation}
In a rank-one theory, Wick-ordered BRST operator reads \cite{BatFra88}
$$
\Omega=T^\dagger_\alpha C^\alpha+C^{\dagger\alpha}T_\alpha+
$$
\begin{equation}
+\left(\frac{1}{2}\bar{C}^\dagger_\gamma U^{\dagger\gamma}_{\alpha\beta}
C^\alpha C^\beta+\frac{1}{2}C^{\dagger\beta}C^{\dagger\alpha}
U^\gamma_{\alpha\beta}\bar{C}_\gamma+C^{\dagger\alpha}
\bar{U}^\gamma_{\alpha\beta}\bar{C}_\gamma C^\beta+
C^{\dagger\beta}\bar{C}^\dagger_\gamma\bar{U}^{\dagger\gamma}_{\alpha\beta}
C^\alpha\right)(-1)^{\varepsilon_\beta}.\label{5.9}
\end{equation}
Original constraint algebra is given by the involution relations
\cite{BatFra88}
\begin{equation}\label{5.10}
[T_\alpha,T_\beta]=U^\gamma_{\alpha\beta}T_\gamma,\quad
[T^\dagger_\beta,T^\dagger_\alpha]=T^\dagger_\gamma
U^{\dagger\gamma}_{\alpha\beta},
\end{equation}
\begin{equation}\label{5.11}
[T_\alpha,T^\dagger_\beta]=\bar{U}^\gamma_{\alpha\beta}T_\gamma+
T^\dagger_\gamma\bar{U}^{\dagger\gamma}_{\beta\alpha}+
\bar{U}^\gamma_{\alpha\delta}\bar{U}^{\dagger\delta}_{\beta\gamma}
(-1)^{\varepsilon_\gamma\varepsilon_\delta}.
\end{equation}

BRST-invariant constraints are
\begin{equation}\label{5.12}
{\cal T}_\alpha=[\bar{C}_\alpha,\Omega]=T_\alpha+
C^{\dagger\beta}U^\gamma_{\beta\alpha}\bar{C}_\gamma(-1)^{\varepsilon_\alpha}
+\bar{U}^\gamma_{\alpha\beta}\bar{C}_\gamma C^\beta(-1)^{\varepsilon_\beta}+
\bar{C}^\dagger_\gamma\bar{U}^{\dagger\gamma}_{\beta\alpha}
C^\beta(-1)^{\varepsilon_\alpha},
\end{equation}
\begin{equation}\label{5.13}
{\cal T}^\dagger_\alpha=[\Omega,\bar{C}^\dagger_\alpha]=T^\dagger_\alpha+
\bar{C}^\dagger_\gamma U^{\dagger\gamma}_{\beta\alpha}C^\beta
(-1)^{\varepsilon_\alpha}
+C^{\dagger\beta}\bar{C}^\dagger_\gamma\bar{U}^{\dagger\gamma}_{\alpha\beta}
(-1)^{\varepsilon_\beta}+
C^{\dagger\beta}\bar{U}^\gamma_{\beta\alpha}\bar{C}_\gamma
(-1)^{\varepsilon_\alpha}.
\end{equation}
Nonzero relations of their commutator algebra read
\begin{equation}\label{5.14}
[{\cal T}_\alpha,{\cal T}_\beta]=U^\gamma_{\alpha\beta}{\cal T}_\gamma+
[\Omega,U^\gamma_{\alpha\beta}]\bar{C}_\gamma
(-1)^{\varepsilon_\alpha+\varepsilon_\beta},
\end{equation}
\begin{equation}\label{5.15}
[{\cal T}^\dagger_\beta,{\cal T}^\dagger_\alpha]=
{\cal T}^\dagger_\gamma U^{\dagger\gamma}_{\alpha\beta}+
\bar{C}^\dagger_\gamma[U^{\dagger\gamma}_{\alpha\beta},\Omega]
(-1)^{\varepsilon_\alpha+\varepsilon_\beta},
\end{equation}
\begin{equation}\label{5.16}
[{\cal T}_\alpha,{\cal T}^\dagger_\beta]=
\bar{U}^\gamma_{\alpha\beta}{\cal T}_\gamma +
{\cal T}^\dagger_\gamma\bar{U}^{\dagger\gamma}_{\beta\alpha}+
[\Omega,\bar{U}^\gamma_{\alpha\beta}]\bar{C}_\gamma
(-1)^{\varepsilon_\alpha+\varepsilon_\beta}+
\bar{C}^\dagger_\gamma[\bar{U}^{\dagger\gamma}_{\beta\alpha},\Omega]
(-1)^{\varepsilon_\alpha+\varepsilon_\beta},
\end{equation}
\begin{equation}\label{5.17}
[\bar{C}_\alpha,{\cal T}_\beta]=
U^\gamma_{\alpha\beta}\bar{C}_\gamma(-1)^{\varepsilon_\beta},\quad
[{\cal T}^\dagger_\beta,\bar{C}^\dagger_\alpha]=
\bar{C}^\dagger_\gamma U^{\dagger\gamma}_{\alpha\beta}
(-1)^{\varepsilon_\beta},
\end{equation}
\begin{equation}\label{5.18}
[{\cal T}_\alpha,\bar{C}^\dagger_\beta]=
\bar{U}^\gamma_{\alpha\beta}\bar{C}_\gamma(-1)^{\varepsilon_\beta}+
\bar{C}^\dagger_\gamma\bar{U}^{\dagger\gamma}_{\beta\alpha}
(-1)^{\varepsilon_\alpha},
\end{equation}
\begin{equation}\label{5.19}
[\bar{C}_\beta,{\cal T}^\dagger_\alpha]=
\bar{C}^\dagger_\gamma\bar{U}^{\dagger\gamma}_{\alpha\beta}
(-1)^{\varepsilon_\beta}+
\bar{U}^\gamma_{\beta\alpha}\bar{C}_\gamma(-1)^{\varepsilon_\alpha}.
\end{equation}
The same as in the case of Weyl-ordered ghost sector, we see that the
extension, represented by the third term in r.h.s. of (\ref{5.11}), is absent
in (\ref{5.16}), although we have in (\ref{5.16}) an admixture of ghost
momenta $\bar{C}_\alpha$ and $\bar{C}^\dagger_\alpha$, instead.

The antibracket algebra, corresponding to (\ref{5.14}) - (\ref{5.19}), reads
\begin{equation}\label{5.20}
({\cal T}_\alpha,{\cal T}_\beta)_\Omega=0,\quad
({\cal T}^\dagger_\alpha,{\cal T}^\dagger_\beta)_\Omega=0,\quad
({\cal T}_\alpha,{\cal T}^\dagger_\beta)_\Omega=0,
\end{equation}
\begin{equation}\label{5.21}
(\bar{C}_\alpha,\bar{C}_\beta)_\Omega=
U^\gamma_{\alpha\beta}\bar{C}_\gamma,\quad
(\bar{C}^\dagger_\beta,\bar{C}^\dagger_\alpha)_\Omega=
\bar{C}^\dagger_\gamma U^{\dagger\gamma}_{\alpha\beta},
\end{equation}
\begin{equation}\label{5.21a}
(\bar{C}_\alpha,\bar{C}^\dagger_\beta)_\Omega=
\bar{U}^\gamma_{\alpha\beta}\bar{C}_\gamma(-1)^{\varepsilon_\beta}+
\bar{C}^\dagger_\gamma
\bar{U}^{\dagger\gamma}_{\beta\alpha}(-1)^{\varepsilon_\alpha},
\end{equation}
\begin{equation}\label{5.22}
(\bar{C}_\alpha,{\cal T}_\beta)_\Omega=\frac{1}{2}[{\cal T}_\alpha,
{\cal T}_\beta],\quad
({\cal T}^\dagger_\beta,\bar{C}^\dagger_\alpha)_\Omega=\frac{1}{2}
[{\cal T}^\dagger_\beta,{\cal T}^\dagger_\alpha],
\end{equation}
\begin{equation}\label{5.23}
(\bar{C}_\alpha,{\cal T}^\dagger_\beta)_\Omega=\frac{1}{2}[{\cal T}_\alpha,
{\cal T}^\dagger_\beta],\quad
({\cal T}_\beta,\bar{C}^\dagger_\alpha)_\Omega=\frac{1}{2}
[{\cal T}_\beta,{\cal T}^\dagger_\alpha],
\end{equation}

Explicit formulas given in Subsections 6.1 and 6.2 demonstrate in a
transparent way that there exists an obvious dualism, represented via the
general correspondence
\begin{equation}\label{5.24}
{\cal T},\quad \bar{\cal P},\quad [\,,\,]\quad\quad\Longleftrightarrow
\quad\quad\bar{\cal P},\quad{\cal T},\quad(\,,\,)_\Omega\quad,
\end{equation}
between the two alternative forms of BRST-invariant constraint algebra.

\section{Conclusion}

In previous sections, we have formulated a new approach to
quantization of gauge-invariant dynamical systems, which is based
substantially on the concept of BRST-invariant constraints.

The hearth of the construction is the new nilpotent ``BRST-charge''
$\Sigma_1$, which lives in yet more extended phase space. Former extended
phase space, spanned by initial phase variables and ordinary ghosts, now
becomes a new ``initial'' space. New canonical pairs $(C^A,\bar{\cal P}_A)$
(\ref{3.34}), (\ref{3.10}) play the role of new ``minimal'' ghosts, while a
new quantum number, the degree, plays the role of a new ghost number.
Regarding these new canonical pairs as ``minimal'' ghosts in effect, we can
introduce new antighosts, $({\cal P}^A,\bar{C}_A)$,
$\varepsilon({\cal P}^A)=\varepsilon(\bar{C}_A)=\varepsilon_A+1$,
$\hbox{gh}({\cal P}^A)=-\hbox{gh}(\bar{C}_A)=\{1;2\}$,
$\hbox{deg}({\cal P}^A)=-\hbox{deg}(\bar{C}_A)=\{1;1\}$,
and Lagrange multipliers, $(\lambda^A,\pi_A)$,
$\varepsilon(\lambda^A)=\varepsilon(\pi_A)=\varepsilon_A$,
$\hbox{gh}(\lambda^A)=-\hbox{gh}(\pi_A)=\{0;1\}$,
$\hbox{deg}(\lambda^A)=-\hbox{deg}(\pi_A)=\{0;0\}$, with the only
nonzero commutators
\begin{equation}\label{6.1}
[{\cal P}^A,\bar{C}_B]=i\hbar\delta^A_B,\quad
[\lambda^A,\pi_B]=i\hbar\delta^A_B,
\end{equation}
and then construct, in a usual way, a new gauge-fixed unitarizing
Hamiltonian.

To realize the above program, we have to construct first a ``minimal''
Hamiltonian $\Xi$, which satisfies the equations
\begin{equation}\label{6.2}
[\Sigma_1,\Xi]=0,
\end{equation}
\begin{equation}\label{6.3}
\varepsilon(\Xi)=0,\quad\hbox{gh}(\Xi)=0,\quad\hbox{deg}(\Xi)=0,
\end{equation}
and boundary conditions
\begin{equation}\label{6.4}
\Xi={\cal H}+\ldots,\quad[{\cal H},\Omega]=0.
\end{equation}
Then, we construct a complete unitarizing Hamiltonian in the standard form,
\begin{equation}\label{6.5}
H=\Xi+(i\hbar)^{-1}[\Sigma,\Psi],
\end{equation}
\begin{equation}\label{6.6}
\varepsilon(\Psi)=1,\quad\hbox{gh}(\Psi)=-1,\quad\hbox{deg}(\Psi)=-1,
\end{equation}
where
\begin{equation}\label{6.7}
\Sigma=\Sigma_1+\pi_A{\cal P}^A,\quad\Psi=\bar{C}_A\chi^A+
\bar{\cal P}_A\lambda^A,
\end{equation}
and $\chi^A$ are gauge-fixing operators. Original Hamiltonian and first-class
constraints are contained in ${\cal H}$ and $\Omega$, respectively, in their
lowest-order terms, when expanded in power series in ordinary ghost operators
$(C^\alpha,\bar{\cal P}_\alpha)$.

Physical observables commute with $\Sigma$, while physical states
are annihilated by this operator. Being a physical scalar product defined
appropriately, physical matrix elements of physical operators are expected to
be gauge independent. If so, one can transit to the unitary limit by choosing
a unitary gauge of the form
\begin{equation}\label{6.8}
\chi^A=0,\quad\chi^A\equiv\{\chi^\alpha;C^\alpha\},
\end{equation}
where $\chi^\alpha$ is an ordinary gauge with respect to original constraints
$T_\alpha$, to identify physical transition amplitude ($S$-matrix) with the
one in the standard BRST-BFV approach. However, when using general
relativistic gauges, the formalism generalizes essentially the standard one
by supporting yet more explicit BRST symmetry of the gauge algebra generating
mechanism.

We finish with the following remark.
Let us consider the standard form \cite{BatVil77}, \cite{FraFra78} of an
unitarizing Hamiltonian in BRST-BFV approach,
\begin{equation}\label{6.9}
H={\cal H}+(i\hbar)^{-1}[Q,\Psi],
\end{equation}
where ${\cal H}$ is a minimal Hamiltonian,
\begin{equation}\label{6.10}
Q=\Omega+\pi_\alpha{\cal P}^\alpha,\quad\Psi=\bar{C}_\alpha\chi^\alpha+
\bar{\cal P}_\alpha\lambda^\alpha,
\end{equation}
\begin{equation}\label{6.11}
[\Omega,\Omega]=0,\quad[{\cal H},\Omega]=0,
\end{equation}
$\Omega$ is a minimal BRST operator, and $\Psi$ is a gauge-fixing Fermion.

We have
\begin{equation}\label{6.12}
H={\cal H}+(i\hbar)^{-1}[\Omega,\bar{\cal P}_\alpha]\lambda^\alpha+
\bar{\cal P}_\alpha{\cal P}^\alpha+\pi_\alpha\chi^\alpha+
\bar{C}_\alpha(i\hbar)^{-1}[\chi^\alpha,\Omega].
\end{equation}
In the second and third terms in r.h.s. we recognize the constraints
${\cal T}_\alpha$ and $X_\alpha$ in their simplest possible form,
\begin{equation}\label{6.13}
{\cal T}_\alpha=(i\hbar)^{-1}[\Omega,\bar{\cal P}_\alpha],\quad
-X_\alpha=\bar{\cal P}_\alpha,
\end{equation}
with $\lambda^\alpha$ and $-{\cal P}^\alpha$ being their respective
Lagrange multipliers.

Then, the fourth and fifth terms are gauge-fixing ones with $\chi^\alpha$ and
$(i\hbar)^{-1}[\chi^\alpha,\Omega]$ being gauge-fixing operators to
${\cal T}_\alpha$ and $X_\alpha$, respectively, and $\pi_\alpha$,
$\bar{C}_\alpha$ being their respective Lagrange multipliers.

So, it appears that the standard Hamiltonian (\ref{6.9}) - (\ref{6.10}) is,
actually, constructed just in terms of the ``standard'' BRST-invariant
constraints (\ref{6.13}) and their respective gauge-fixing operators.

However, as compared with general operators $T_A$, which satisfy (\ref{14}),
(\ref{17}) only, the constraints (\ref{6.13}) are rather special ones in the
sense that they relate to a special basis in terms of $T_A$.

Contrary to that, our new Hamiltonian (\ref{6.12}), living in yet more
extended phase space, is constructed directly in terms of general operators
$T_A$ subject to (\ref{14}), (\ref{17}) only. Thus, the involvement of the new
variables $(C^A,\bar{\cal P}_A)$, $({\cal P}^A,\bar{C}_A)$,
$(\lambda^A,\pi_A)$ is just a price of arbitrariness in choosing possible
basis to the general BRST-invariant constraints $T_A$.

It is also worthy to mention that the step, we have made from (\ref{6.9}) -
(\ref{6.11}) to (\ref{6.2}) - (\ref{6.7}), seems to be only the first one in,
possibly infinite, hierarchy of Hamiltonians.

{\bf Acknowledgements}

The authors are grateful to Maxim Grigoriev and Alexei Semikhatov for
fruitful discussions. I.A.B. would like to thank Robert Marnelius for
illuminating discussions of aspects related to the quantum antibrackets and
operator-valued master equation. The work of I.A. Batalin is supported in
part by the President grant 00-15-96566, the RFBR grant 02-01-00930, and the
INTAS grant 00-00262. The work of I.V. Tyutin is supported in
part by the President grant 00-15-96566, the RFBR grant 02-02-16944, and the
INTAS grant 00-00262.

\end{document}